\documentclass[12pt]{iopart}
\usepackage{graphicx}
\usepackage{wrapfig}
\expandafter\let\csname equation*\endcsname\relax
\expandafter\let\csname endequation*\endcsname\relax
\usepackage{amsmath}
\usepackage{color}
\newcommand{\wsqcm}{\textrm{\,W\,cm}^{-2}}
\newcommand{\eqb}{\begin{eqnarray}}
\newcommand{\eqe}{\end{eqnarray}}
\newcommand{\uperp}{{u_\bot}}
\newcommand{\gperp}{{g_\bot}}
\newcommand{\diff}{{\rm d}}

\newcommand{\eps}{\epsilon}

\newcommand{\apj}{Astrophysical J.}

\newcommand{\pre}{Phys.\ Rev.\ E}

\begin{document}
\title{Pair plasma cushions in the hole-boring scenario}
\author{J.G. Kirk$^1$, A.R. Bell$^{2,3}$, C.P. Ridgers$^{3,4}$}
\address{$^1$ Max-Planck-Institut f\"ur Kernphysik, Heidelberg, Germany}
\address{$^2$ Clarendon Laboratory, University of Oxford, Oxford, UK}
\address{$^3$ Central Laser Facility, STFC Rutherford-Appleton Laboratory, 
Chilton, UK}
\address{$^4$ Department of Physics, University of York, York, UK}
\begin{abstract}
  Pulses from a 10~PW laser are predicted to produce large numbers of
  gamma-rays and electron-positron pairs on hitting a solid target.  
  However, a pair plasma, if it accumulates in front of the target, 
may partially shield it from the pulse.  Using  
stationary, one-dimensional solutions of the two-fluid
  (electron-positron) and Maxwell equations, including a classical radiation
  reaction term, we examine this effect in the hole-boring scenario. We 
find the collective effects of a pair plasma 
\lq\lq cushion\rq\rq\ substantially reduce the 
  reflectivity, converting the absorbed flux into high-energy gamma-rays.
There is also a modest increase in the laser intensity needed to achieve
threshold for a non-linear pair cascade.  
\end{abstract}

Accepted for publication in \PPCF

\section{Introduction}
Particle-in-cell simulations of ultra-intense laser pulses interacting
with plasma now probe the regime in which non-linear QED processes are
important \cite{nerushetal11,ridgersetal12,bradyetal12}, and predict the
production of large numbers of gamma-rays and electron-positron pairs
when the laser interacts with either an over-dense or an under-dense plasma.
Next generation lasers (10~PW) will be able to test these predictions.
However, it is still not clear whether or not the fully non-linear pair cascade 
predicted by Bell \& Kirk~\cite{bellkirk08} and 
Fedotov et al~\cite{fedotovetal10} will be achieved. 
For counter-propagating pulses, the threshold is expected to lie below
a single pulse intensity of $10^{24}\wsqcm$. But 
simulations of interactions with over-dense plasmas \cite{ridgersetal12},
which are the more straightforward experimental set-up, have
identified several effects that might raise the required threshold
intensity.

One effect with similar consequences that has so far not been analyzed
is the screening by a cloud or \lq\lq cushion\rq\rq\ of pair plasma in
the laser pulse just ahead of the target. As this cushion approaches
the critical density, collective effects in the pair plasma 
can be expected to slow down the laser pulse and reflect or
absorb it. Such cushions are observed in PIC simulations of linearly 
polarized laser pulses 
interacting with dense, solid targets (see, for example, 
figure~1 in \cite{ridgersetal12}), but their dynamics 
are complex. In this paper we do not attempt an analysis of 
simulation results. Instead, we try 
to gain a qualitative understanding of
pair cushions by investigating stationary solutions in 
a highly simplified situation. Although they might be difficult 
to realize in practice, these solutions provide an easily quantifiable
framework in which to interpret PIC simulations and discuss experimental 
set-ups.  

\begin{figure}
\hfill\input{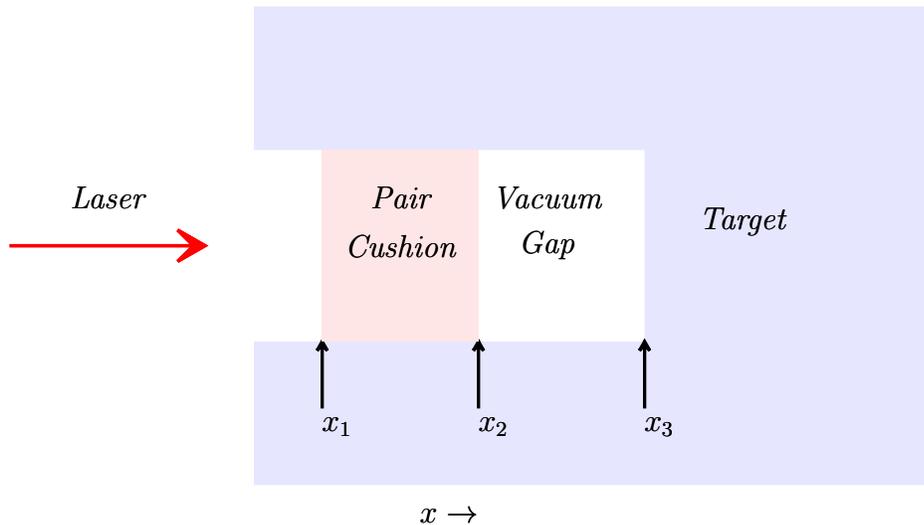}
\caption{\label{sketch}%
Sketch of the idealized hole-boring scenario. 
A pair cushion is located in the laser beam at
$x_1<x<x_2$. Note that a 
vacuum gap, at $x_2<x<x_3$, 
must separate the cushion from the hole-boring front.
}
\end{figure}

We consider the interaction of a circularly polarized laser beam with
a solid target in the hole-boring scenario
\cite{robinsonetal09,schlegeletal09,naumovaetal09}, as sketched in Fig.~\ref{sketch}. 
A pair plasma cushion, located in the region
excavated by the beam, is described by a one-dimensional, cold, 
two-fluid model that includes a classical radiation reaction term.  
Stationary solutions are found and matched to the boundary conditions
of the incoming laser beam on one side, and the standing wave in the
vacuum gap, on the other. Section~\ref{holeboring} describes the 
hole-boring scenario and, in particular, the relativistic dynamics 
of the hole-boring front; 
section~\ref{twofluidmodel} presents the two-fluid 
equations and the method of their solution in the rest frame of the hole-boring
front; section~\ref{results} analyzes the properties of these solutions and
section~\ref{applications} discusses a practical application. We conclude 
with a discussion of the physical significance of the solutions.

\section{The hole-boring scenario}
\label{holeboring}
\subsection{Dynamics of the hole-boring front}
The one-dimensional model of hole-boring by a circularly polarized
plane wave is based on the properties of the \lq\lq hole-boring
front\rq\rq\cite{robinsonetal09,schlegeletal09,naumovaetal09} 
that divides the vacuum waves on one
side from the high density plasma (originally solid) on the other. The
front reflects the ions and electrons that stream into it as it
advances into the solid, by means of a charge-separated region that
supports a strong, longitudinal electric field. At the same time, the
front perfectly reflects the circularly polarized laser that is
incident on its other surface.  A closer look reveals that electrons
are absent in the charge-separated region, which is bounded on the
target side by an \lq\lq electron sheath\rq\rq, a thin enhancement in
the electron density at which the laser is reflected.  The target ions
stream in through the electron sheath, are brought to rest by the
longitudinal electrostatic field, and subsequently accelerated back
through the electron sheath \cite{schlegeletal09}. 
The key parameter in this scenario is 
the ratio of the 
incident laser intensity $I^+$, assumed be a circularly polarized, 
monochromatic plane wave of (angular) frequency $\omega_{\rm lab}$
propagating in the positive $x$ direction,  
and the solid target density
$\rho$ (both measured in the lab.\ frame), which are combined into the 
dimensionless quantity
\eqb
X&=& I^+/\left(\rho c^3\right)
\label{xdefinition}
\\
&=&0.37\times \left(\frac{I^+}{10^{24}\wsqcm}\right)
\left(\frac{\rho}{1\,\textrm{g\,cm}^{-3}}\right)^{-1}
\nonumber
\eqe
For a laser pulse of intensity 
$I^+<10^{24}\wsqcm$ impacting a metal target, $X\ll1$.
The thickness of the
charge-separated region is then
approximately $\sqrt{X}c/3\omega_{\rm i}$ \cite{schlegeletal09}, where
$\omega_{\rm i}$ is the ion plasma frequency 
in the fully ionized target.
This length is small
compared to the laser wavelength, which is 
the characteristic dimension of the pair cushion. Therefore,
we will treat the hole-boring front as a singular surface 
current --- a perfect mirror, at which the electric field vanishes. 

The speed of advance of the front
into the solid, as well as the energy in the laboratory frame of the
reflected ions, is given by equating the pressure exerted by the laser
with that exerted by the in-streaming and reflected ions, neglecting
the electron inertia. 
Allowing for a reflected wave of intensity $I^-$, the energy-momentum tensor 
$T^{\mu\nu}$ of the radiation field has the following non-vanishing elements
\eqb
T^{00}\,=\,T^{11}\,=\,\left(I^+ + I^-\right)/c\,=\,I^+\left(1 + R\right)/c
\nonumber\\
T^{01}\,=\,T^{10}\,=\,\left(I^+ - I^-\right)/c\,=\,I^+\left(1-R\right)/c
\label{stress}
\eqe
where the reflectivity is $R=I^-/I^+$.
The front is assumed to move into the solid 
(which is at rest in the lab.\ frame) 
at constant speed $c\beta_{\rm f}$, and the elements $T'^{\mu\nu}$ 
of the energy-momentum tensor 
of the laser in the 
rest-frame of the front (the \lq\lq HB-frame\rq\rq) follow from Lorentz 
boosting (\ref{stress}):
\eqb
T'^{01}&=&T'^{10}\,=\,\varGamma_{\rm f}^2\left[\left(1-\beta_{\rm f}\right)^2 - 
\left(1+\beta_{\rm f}\right)^2 R\right]I^+/c
\\
T'^{00}&=&T'^{11}\,=\,\varGamma_{\rm f}^2\left[\left(1-\beta_{\rm f}\right)^2 + 
\left(1+\beta_{\rm f}\right)^2 R\right] I^+/c
\label{hbstress}
\eqe
where $\varGamma_{\rm f}=\left(1-\beta_{\rm f}^2\right)^{-1/2}$.
Thus, the reflectivity in the rest frame of the  hole-boring front 
is
\eqb
R'&=&\left(\frac{1+\beta_{\rm f}}{1-\beta_{\rm f}}\right)^2 R
\\
&=&D^{-4} R
\eqe
where the Doppler factor 
\eqb
D&=&\left(\frac{1-\beta_{\rm f}}{1+\beta_{\rm f}}\right)^{1/2}
\eqe
is the ratio of the laser frequency $\omega$ 
in the HB-frame to its value $\omega_{\rm lab}$ in the 
lab.\ frame.

On the other hand, the elements of the energy-momentum 
tensor on the target side 
of the hole-boring front are found by assuming perfect reflection of the ions.
In the HB-frame, therefore, two mono-energetic beams of velocity 
$\pm c\beta_{\rm f}$ and proper density $\rho$ exist immediately behind 
the front, so that
\eqb
T'^{00}&=& 2\rho c^2\varGamma_{\rm f}^2
\nonumber\\
T'^{11}&=& 2\rho c^2 \beta_{\rm f}^2\varGamma_{\rm f}^2\\
T'^{01}&=&0
\eqe
The requirement that $T'^{11}$ be continuous across the front determines
the advance speed, provided $R'$ is known:
\eqb
\beta_{\rm f}&=&\sqrt{\xi}/\left(1+\sqrt{\xi}\right)
\label{generaladvance}
\eqe
where the reflection-modified $X$ parameter is defined as
\eqb
\xi&=&X(1+R')/2
\label{xidefinition}
\eqe
The Doppler factor and (dimensionless) $x$-component of the 
four-velocity of the front, 
$u_{\rm f}=\beta_{\rm f}\varGamma_{\rm f}$, are also functions
of $\xi$ alone:
\eqb
D&=&\left(1+2\sqrt{\xi}\right)^{-1/2}
\\
u_{\rm f}&=&D\sqrt{\xi}
\eqe
and the reflectivity in the lab.\ frame is
\eqb
R&=&\left(1+2\sqrt{\xi}\right)^{-2}R'
\eqe

In the standard hole-boring scenario, perfect reflection is assumed in
the rest frame of the hole-boring front: $R'=1$, $\xi=X$, in which case
these expressions agree with those given by \cite{robinsonetal09}. In
this case, the electric and magnetic fields, as seen in the
HB-frame, form a standing wave. They 
are everywhere parallel, and lie in a plane that contains
the $x$-axis and rotates about it. The field  magnitudes are
constant in time at each position, but at a given instant
vary sinusoidally in $x$ with
equal amplitudes 
\eqb 
E_{\rm ampl}&=&2D\sqrt{4\pi I^+/c}
\label{amplitude}
\eqe 
and a phase difference of $\pi/2$. As we show below, a pair cushion
greatly reduces $R'$, leading to a smaller speed of advance of the
hole-boring front, and a reduced electric field amplitude.

\section{Two-fluid model of the pair cushion}
\label{twofluidmodel}
\subsection{Governing equations}
In the presence of a pair plasma, the incident and reflected 
waves in the excavated channel are strongly 
coupled, and the relevant solutions are not vacuum waves, but 
nonlinear, transverse 
electromagnetic modes
of superluminal phase speed. 
We use a cold, two-fluid (electron and positron) 
description to analyze these waves. The continuous
charge and current distributions
are related to the fields by Maxwell's equations. The fluids
obey equations of motion that contain not
only the wave fields, but also the classical radiation reaction force,
thus taking into account the discrete nature of the fluid constituents.
The cartesian 
four-velocity components and 
the proper densities are denoted by $u^\pm_{x,y,z}$ and $n^\pm$. Since we 
treat circular polarization, it is convenient 
to use rotating coordinates:
\eqb
\uperp^\pm&=&u^\pm_y+i u^\pm_z
\qquad
E\,=\,E_y+iE_z\qquad B\,=\,B_y+iB_z
\eqe

In order to find nonlinear solutions that are homogeneous in the 
$y$--$z$ plane, we make a number of simplifications: 
Firstly, in the transverse electromagnetic waves of interest here, electrons
and positrons have the same density and oppositely directed transverse
momenta: $n^-=n^+=n$ and $\uperp^-=-\uperp^+=\uperp$. It follows that 
$E_x=0$. Secondly, we 
look for solutions which the 
fluids do not stream along $x$ in the frame in which the 
hole-boring front is stationary: $u^\pm_x=0$.
In the following, the $\pm$ 
notation is dropped and the equations
presented apply to the electron fluid in this frame. 

The classical radiation reaction force in the Lorentz-Abraham-Dirac 
formulation is: 
\eqb
g^\mu&=&\frac{2e^2}{3mc^3}\left(\frac{\diff^2 u^\mu}{\diff \tau^2}
- u^\mu\left|\frac{\diff u^\nu}{\diff\tau}\right|^2\right)
\label{LAD}
\eqe
and it is clear that the spatial components lie in the $y$--$z$ plane when
$u_x\equiv0$. This property is shared by the Landau-Lifshitz formulation
of radiation reaction, in which the derivatives in (\ref{LAD}) are replaced
using the Lorentz equation of motion (see \cite{dipiazzaetal12} for a review).
Thus, the $x$-component of the fluid equation of motion is unaffected by
radiation reaction: 
\eqb
\left(\gamma
\frac{\partial}{\partial t}+c 
u_x\frac{\partial}{\partial x}\right)u_x
&=&-\frac{e}{mc}
\textrm{Im}\left(\uperp B^*\right)
\eqe
where $\gamma=u^0$. Solutions with $u_x=0$ for 
all $x$ and $t$, therefore, 
require the transverse velocity and magnetic field vectors to 
be parallel:
\eqb 
\textrm{Im}\left(\uperp B^*\right)&=&0 
\label{uparallelB}
\eqe

On the other hand, the (complex) equation of motion in the transverse plane 
contains a term due to radiation reaction: 
\eqb 
\gamma\frac{\partial
  \uperp}{\partial t}&=& -\frac{e}{mc}\gamma E + \gperp 
\label{eqmottransverse}
\eqe 
where $\gperp$ is the spatial part of $g^\mu$ in rotating coordinates, and 
we have set $u_x=0$.
For these transverse fields ($E_x=0$, $B_x=0$), the set of governing 
equations is completed by the 
Faraday and Amp\`ere equations:
\eqb 
\frac{\partial E}{\partial x}-\frac{i}{c}\frac{\partial B}{\partial t} &=&0 
\\
\frac{\partial B}{\partial
  x}+\frac{i}{c}\frac{\partial E}{\partial t} &=&i8\pi e n \uperp
\eqe 
and the equation of continuity: 
\eqb
\frac{\partial}{\partial t}\left(\gamma n\right)&=&0
\eqe

\subsection{Method of solution}
We seek solutions that are separable in $x$ and $t$ 
in the HB-frame. 
In particular, for a monochromatic wave
of angular frequency, $\omega$,
the quantities $E$, $B$, and $\uperp$
are proportional to $\textrm{e}^{i\omega t}$, whereas $n$ and
$\left|\uperp\right|$ are constant in time. 
Since force balance along $x$ (\ref{uparallelB}) 
requires the fluid velocity to be parallel to
the magnetic field, the complex 
variables $E$, $B$ and $\uperp$ can 
be replaced by three real, positive, dimensionless 
amplitudes $a$, $b$ and $u$, together
with two phases, $\phi$ and $\delta$, all of which are functions of 
$x$ only:
\eqb
E&=&\left(\frac{mc\omega}{e}\right)a\textrm{e}^{i\phi +i\omega t}\\
B&=&\left(\frac{mc\omega}{e}\right)ib\textrm{e}^{i\phi+i\delta +i\omega t}\\
\uperp&=&iu\textrm{e}^{i\phi+i\delta +i\omega t}
\eqe
Substituting these into (\ref{eqmottransverse}), the transverse equations of 
motion become:
\eqb
u&=&a \cos\delta
\label{eqyz1}\\
\delta&=&\arctan\left(\eps u^3\right)
\label{eqyz2}
\eqe
where 
\eqb
\eps&=&\frac{2}{3}\frac{\omega e^2}{mc^3}
\\
&=&1.18\times10^{-8}\,D\,\lambda_{\mu{\rm m}}^{-1}
\eqe
with $\lambda_{\mu{\rm m}}$ the laser (vacuum) wavelength in the lab.\ frame 
in microns. 
The Lorentz-Abraham-Dirac 
form (\ref{LAD}) of the radiation reaction term was 
used in deriving (\ref{eqyz2}), and only the leading contribution
in an expansion
in $1/\gamma$ was retained:
\eqb
\gperp&\approx&
-\left(2e^2\omega/3mc^3\right)\omega\gamma^4\uperp
\label{gperpapprox}
\eqe
The Landau-Lifshitz form yields
exactly the same result in this limit.   

The pair fluids do not contribute to the 
$(1,1)$ and $(0,1)$ components of the energy-momentum tensor. The latter is,
therefore, just the Poynting flux:
\eqb
T'^{01}&=& \left(\frac{m^2c^2\omega^2}{8\pi e^2}\right)P
\nonumber\\
P&=& 2ab\cos\delta
\label{poyntingexpr}
\eqe
and the former is the energy-density of the fields:
\eqb
T'^{11}&=&\left(\frac{m^2c^2\omega^2}{8\pi e^2}\right)U
\nonumber\\
U&=&a^2+b^2
\eqe 
The Faraday and Amp\`ere equations take the form:
\eqb
\diff a/\diff x&=&b\sin\delta \quad 
\diff \phi/\diff x\,=\,-\left(b/a\right)\cos\delta
\label{faraday2}\\
\diff b/\diff x&=&-a\sin\delta \quad 
\diff \delta/\diff x\,=\,
\left[\left(n-n_{\rm cr}\right)/n_{\rm cr}\right]\left(a/b\right)\cos\delta 
-\diff \phi/\diff x
\label{ampere2}
\eqe
and 
can be used to evaluate the 
divergence (in this case, derivative with respect to $x$) of
$T'^{01}$ and $T'^{11}$:
\eqb
\frac{\diff P}{\diff x} &=&
-2\left(n/n_{\rm cr}\right)a^2\sin\delta\cos\delta
\label{energyconsv2}
\\
\frac{\diff U}{\diff x}&=&0
\label{momconsv2}
\eqe
where $n_{\rm cr}=m\omega^2/8\pi e^2$ is the critical proper density. 

The solution of the 
system (\ref{eqyz1}), (\ref{eqyz2}), (\ref{faraday2}) and (\ref{ampere2})
can be reduced to a quadrature:
\eqb
x&=&\int\,\diff \delta
\frac{4-3\cos^2\delta}{3\sin^2\delta\cos\delta}
\left[
  \left(\frac{r_{\rm c}\cos^4\delta}{\sin\delta}\right)^{2/3} - 1\right]^{-1/2}
\label{deltaprimeeq2}
\eqe 
where
\eqb 
r_{\rm c}&=&\eps U^{3/2}\,=\,\textrm{constant}
\label{rcdef} 
\eqe
Having found $\delta(x)$, $a$ and $u$ follow from (\ref{eqyz1})
and (\ref{eqyz2}). The constant pressure condition, 
(\ref{momconsv2}), determines the magnetic field 
$b$ in terms of the 
constant $a_0=\sqrt{a^2+b^2}=\sqrt{U}$ and the density 
 follows from energy conservation, (\ref{energyconsv2}):
\eqb
\frac{n}{n_{\rm cr}}&=&
\frac{2a^2+3a^2\sin^2\delta-a_0^2}{a^2\left(4-3\cos^2\delta\right)}
\label{densityeq}
\eqe
Finally, the phase $\phi$ is evaluated from
\eqb
\phi&=&-\int \diff a\frac{\tan(\delta)}{a}
\label{phiintegral}
\eqe

This solution depends on only one parameter,
$r_{\rm c}$, which determines the importance of radiation 
reaction, and is 
equivalent to $R_{\rm C}$ in the notation of 
de~Piazza~et~al~\cite{dipiazzaetal12}. 
The \lq\lq classical radiation-dominated regime\rq\rq\ 
corresponds to $r_{\rm c}>1$.  
In terms of laser parameters,
\eqb
r_{\rm c}
&=&7.4 \lambda_{\mu{\rm m}}^2\left(I^+/10^{24}\wsqcm\right)^{3/2}
D\left(1+R'\right)^{3/2}
\nonumber
\eqe

\begin{figure}
\hfill\input{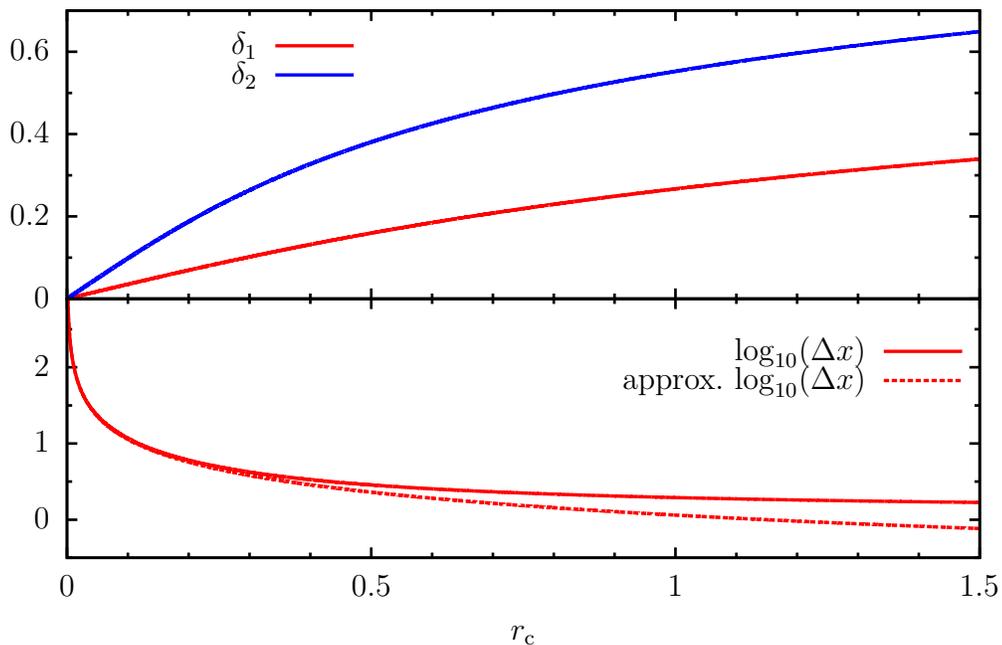}
\caption{\label{endpoint}%
Top: the phase-shift of the magnetic field 
caused by radiation reaction at the upstream ($\delta_1$) 
and downstream ($\delta_2$) edges of the pair cushion (the angle between the 
electric and magnetic field vectors is $\pi/2+\delta$).
Bottom: the maximum thickness $\Delta x=x_2-x_{\rm min}$ 
of the cushion (in units of $c/\omega$) compared
to the approximate expression (\ref{xminapprox}),
both as functions of the radiation reaction parameter
$r_{\rm c}$ defined in (\ref{rcdef})}
\end{figure}

\subsection{Boundary conditions}
At the downstream boundary, $x=x_2$, (see Fig.~\ref{sketch}) 
the fields of the pair cushion 
must match those of either the hole-boring
front or a vacuum gap. 
Continuity of the energy-momentum
tensor components $T'^{01}$ requires zero Poynting flux, because the
hole-boring front is assumed perfectly reflecting in the HB-frame. 
Using the notation $a_{1,2}=a\left(x_{1,2}\right)$ etc., 
this implies either $b_2=0$, or
$a_2=u_2=\delta_2=0$. 
The latter possibility is, however, unphysical, since it
implies $\diff a/\diff x>0$ for $x\rightarrow x_2$, 
leading to a negative value of
$a$ just upstream of this boundary. 
Thus, $b_2=0$, $a_2=a_0$, and 
$\delta_2$ follows from (\ref{eqyz1}) and (\ref{eqyz2}).  
This means that the edge of the pair cushion cannot be located at the 
mirror, where continuity of the tangential component of $E$ requires 
it to vanish.
Because the finite-density cushion cannot carry a 
singular current sheet at the surface $x=x_2$, the
transverse component of $B$ must also be continuous across
it. Therefore, this point lies at a node of the magnetic field not only of
the plasma wave, but also of the standing wave that occupies the
vacuum region $x>x_2$.
The cushion must, therefore, be separated from the mirror 
by a vacuum gap of thickness
$(j+1/2)\pi c/\omega$, where $j=1,2,\dots$.
According to (\ref{densityeq}), the pair
density reaches the critical value at the edge of this gap: 
$n_2=n_{\rm cr}$.

\begin{figure}
\hfill\input{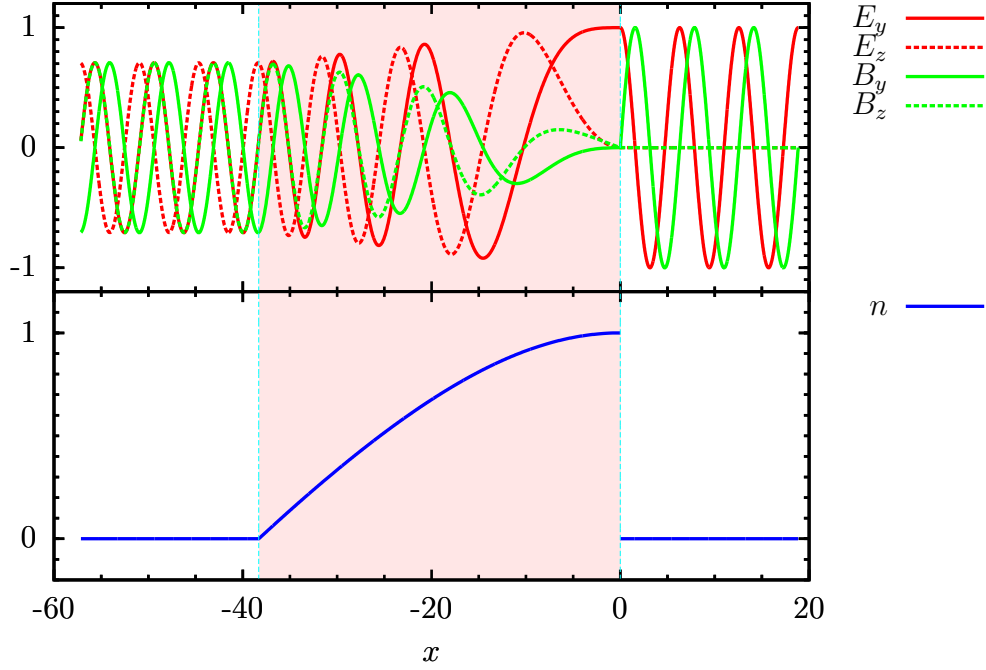}
\caption{\label{profile}%
Top: the spatial profile of the electromagnetic fields, as seen in the 
rest frame of the hole-boring front at $t=0$, normalized
to the value $a_0mc\omega/e$.
Bottom: the proper density, normalized to the critical density 
$\omega^2 m/\left(8\pi e^2\right)$. 
The edges of the pair front are indicated by vertical lines
at $x_1=-38.3$ and $x_2=0$. Note that the 
fields are continuous, but the current-density discontinuity at $x=0$, 
imposes a discontinuity on the  
$x$ derivative of $B$ (but not of $E$). In this figure, 
the radiation reaction parameter is $r_{\rm c}=0.03$.}
\end{figure}

At the upstream boundary, $x=x_1$, the fields in the cushion
must match the vacuum fields of the incident and reflected laser beams.
The location of this point is fixed 
by the number of pairs contained per unit area of the cushion, which 
rises with increasing cushion thickness 
from zero when $x_1=x_2$ to a maximum value determined
by the point at which the proper pair density vanishes. 
The electric field amplitude $a=a_{\rm min}$ at which this occurs
can be found from (\ref{eqyz1}), (\ref{eqyz2}) and (\ref{densityeq}):
\eqb
\eps^2\left(5 a_{\rm min}^2-a_0^2\right)^4+54a_{\rm min}^2-27a_0^2&=&0
\eqe
The corresponding position   
sets the maximum thickness, $\Delta x$ (in units of $c/\omega$), 
of the pair front
compatible with a physically acceptable solution.
In general, a quadrature is needed to find this quantity. However, for 
$r_{\rm c}\ll1$, one finds 
$\delta_2\approx r_{\rm c}$, $\delta_1\approx r_{\rm c}/2^{3/2}$, and 
$a_1\approx a_0/\sqrt{2}$. This leads to the approximate expression:
\eqb
\Delta x&=& \left|x_2-x_1\right|_{\rm max}
\nonumber\\
&\approx&\frac{1}{r_{\rm c}}
\int_{1/\sqrt{2}}^1\,\frac{\diff y}{y^3\sqrt{1-y^2}}
\nonumber\\
&\approx&
1.1478/r_{\rm c}
\label{xminapprox}
\eqe
Fig~\ref{endpoint} compares this result to the numerically 
evaluated quadrature.

\section{Results}
\label{results}
The spatial dependence of the electromagnetic fields is shown in 
Fig~\ref{profile}, for $r_{\rm c}=0.03$ at time $t=0$ (the 
fields are proportional
to $\textrm{e}^{i\omega t}$). 
In this example, the pair front has been chosen
to have its maximum thickness, i.e., the density vanishes at the 
upstream edge. It then rises monotonically, reaching the critical value at 
the downstream edge of the front, where $x=0$. 
For $r_{\rm c}\ll1$, the thickness of the front
is large compared to the laser wavelength in vacuo, and it is easy to 
see that the fields belong to an oscillation whose wavelength grows 
as the pair density rises. 
The increase in wavelength corresponds to 
an increase of the 
phase-velocity of the wave $v_{\rm ph}=\omega/k$, 
but a decrease of the group velocity $v_{\rm g}=c^2 k/\omega$, 
which vanishes at the downstream edge. 

Although the complex fields are separable functions of $x$ and $t$, their
real components are not. This means that the solutions are not standing
waves in the strict sense \cite{marburgertooper75}, except in the vacuum
gap between the cushion and the target, where the Poynting flux vanishes. 
However, they can easily be visualized as a stationary structure that 
rotates about the $x$-axis. This structure forms a helix or screw thread of 
variable pitch. Upstream of the cushion, the pitch is almost constant, but 
with a small periodic fluctuation due to 
the finite amplitude of the reflected wave. Inside the cushion, the pitch
increases monotonically with the pair density, becoming infinite 
at the downstream edge, where $n_2=n_{\rm crit}$. 

Vacuum waves propagate 
both upstream and downstream of the pair front. 
Upstream 
(the region $x<38.3$ in Fig.~\ref{profile}), the amplitude of the backward
propagating wave is given by the reflectivity of the overall system consisting 
of pair front plus hole-boring front. For a pair front of maximum thickness,
\eqb
R'&=&\frac{a_0^2-2a_1b_1\cos\delta_1}{a_0^2+2a_1b_1\cos\delta_1}
\label{reflectivity}
\eqe
and, assuming $r_{\rm c}\ll 1$, one finds
\eqb
R'&\approx&\delta_1^2/4\nonumber\\
&\approx&r_{\rm c}^2/32
\eqe
For the parameters of Fig.~\ref{profile}, $R'\approx2.8\times10^{-5}$, 
and the forward propagating wave dominates, so that 
$E_y\approx B_z$ and
$E_z\approx -B_y$. 

Downstream of the pair front, where $x>x_2$, the forward and backward 
propagating waves are of equal magnitude, since the hole-boring front is 
assumed to be perfectly reflecting. Here, the electric 
and magnetic fields are everywhere parallel. At $t=0$, they are directed 
along the $y$ axis, and rotate together in a clockwise sense, when viewed along
the positive $x$-direction.

At both the upstream and downstream edges, the electromagnetic fields
are continuous, as are the fluxes of energy, $T'^{01}$, and
$x$-momentum, $T'^{11}$. The density, however, is discontinuous (note
that $u_x=0$), although only at the downstream edge for a front of
maximum permitted thickness. A discontinuity in the density implies a
discontinuity also in the current density. As a result, the
$x$-derivative of the magnetic field has a discontinuity, but that of
the electric field does not. 

\begin{figure}
\hfill\input{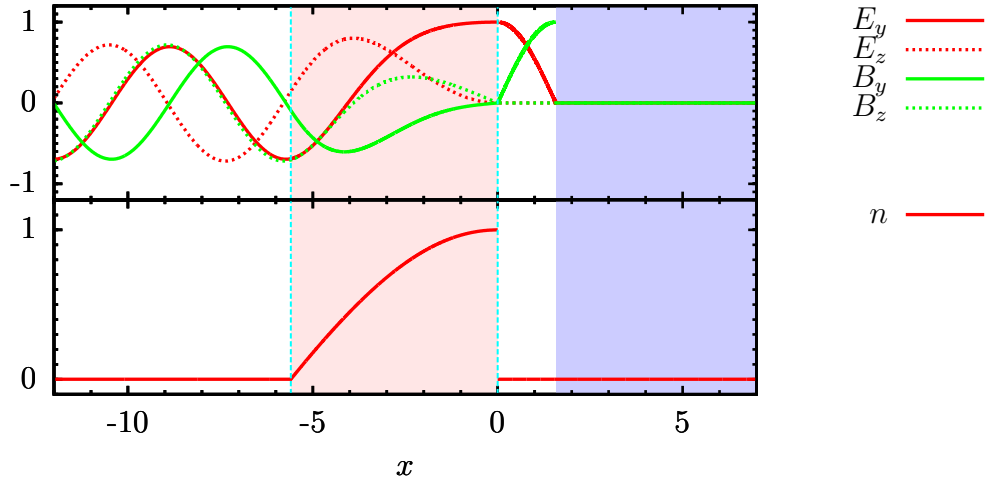}
\caption{\label{fullprofile}%
The field and pair density profiles in the presence of a pair cushion.
(shaded red) for $r_{\rm c}=0.2$. The target is shaded blue. The boundaries of
the pair cushion are at $x_1=-5.6$ and $x_2=0$.
}
\end{figure}

The edge of the pair cushion at $x=x_2$ is located at a node of
the magnetic field in the standing wave that lies downstream of it.
The hole-boring front itself, however, can be located at any of the nodes of
the electric field in this wave. Choosing the minimum vacuum gap size,
the fields in the pair cushion and vacuum gap 
are illustrated in Fig.~\ref{fullprofile}, for $r_{\rm c}=0.2$.

\begin{figure}
\hfill\input{fig5.tex}
\caption{\label{reflect}%
As functions of the laser intensity
$I_{24}\times 10^{24}\wsqcm$, for a wavelength of $1\,\mu\textrm{m}$ and
an aluminium target:\\
{\em Top panel}: 
The reflectivity $R$ in the lab.\ frame and $R'$ in the frame of the 
hole-boring front when a pair front of maximal extent is present. 
For comparison, the 
reflectivity $R_{\rm vac}$ in the lab.\ frame in the absence of a pair front
is also shown.\\
{\em Centre panel}:
The kinetic energy in the lab.\ frame $E_{\rm ion}$ of an ion reflected
into the target by the hole-boring front, the thickness of the pair cushion, 
in units of $c/\omega$, and the parameter 
$r_{\rm c}$ characterizing the strength of classical 
radiation reaction.
\\
{\em Bottom panel}: 
  The non-linear QED parameter reached by electrons and
  positrons in the pair front
for maximal suppression due to absorption of
  the laser by the pair front, $\eta_{\rm pairs}$ 
  and for negligible absorption $\eta_{\rm vac}$.
For comparison, the quantity $\eta_{\rm BK}$,
  as given by \cite{bellkirk08}, is plotted. This applies to
  counter-propagating laser beams, each of the given intensity, in an
  under-dense plasma. 
} 
\end{figure}

\section{Applications}
\label{applications}
When the pair front contains very few particles, $b_1\rightarrow b_2=0$ and the 
reflectivity approaches its maximum value, which is
that of the standard scenario without a pair front,
$R'=1$, $R=1/\left(1+2\sqrt{X}\right)^{-2}$. 
However, the minimum value of $R'$, which is 
attained for the maximum number
of pairs consistent with a stationary solution, 
depends on the radiation reaction parameter 
$r_{\rm c}$, which, in turn, depends not only on 
the hole-boring parameter $X$, but also on the 
laser wavelength and the number of pairs contained in the cushion.

As an example, we consider the effect of a pair cushion 
when an intense laser pulse 
of wavelength $\lambda_{\mu{\rm m}}=1$ impacts an aluminium
target ($\rho=2.7\times10^3\,\textrm{kg\,m}^{-3}$, $Z=13$).
This target is over-dense, since the ratio of the 
electron density to the critical density is 
$698\,\lambda_{\mu{\rm m}}^2$. Therefore, the hole-boring scenario can be 
expected to apply provided the laser is not intense enough to 
to render it relativistically under-dense, which implies the restriction
$I_{24}<1.33\,\lambda_{\mu{\rm m}}^2$, assuming a circularly polarized pulse. 
For laser intensities in this range,
the stationary solution with the largest 
number of pairs is found by iteratively 
solving (\ref{generaladvance}) and (\ref{xidefinition}) 
together with (\ref{reflectivity}).

The top panel of Fig.~\ref{reflect} shows
the reflectivities in the lab.\ and hole-boring frames
assuming a pair front of maximum extent is present in the channel.
At low intensities, the laser is almost completely absorbed 
by the pair cushion. In the approximation 
used to describe radiation reaction, (\ref{gperpapprox}), 
the energy absorbed is converted entirely into transversely directed, 
synchrotron-like radiation. Thus, in the HB-frame the efficiency 
of conversion of laser light into high energy photons is $1-R'$,
i.e., close to $100\,\%$ at low intensities and $87\,\%$ at 
an intensity of
$10^{24}\wsqcm$.

The middle panel of Fig.~\ref{reflect} shows the kinetic 
energy in the lab.\ frame 
of an aluminium ion reflected back into the target off the advancing 
hole-boring front: $E_{\rm ion}=2Mc^2\beta{\rm f}^2\Gamma_{\rm f}^2$.
In the case shown here, the speed of advance of the 
hole-boring front $c\beta_{\rm f}$, remains non-relativistic, even for 
the highest intensity plotted. 
It is assumed here that, for a given laser intensity, 
the pair cushion attains its maximum possible
thickness, which is plotted in units of $c/\omega$ as a function of 
laser intensity in this panel. 
Also shown is the parameter
$r_{\rm c}$ which characterizes the importance 
of the classical radiation reaction force. Above 
$I\approx3\times 10^{23}\wsqcm$, this parameter exceeds unity, i.e., the 
energy radiated by a single electron or positron in one laser period is 
greater than the particle energy.  Radiation 
reaction is responsible for the change in slope at roughly this intensity
of the curves depicting $R'$ and $R$ in the upper panel. When quantum effects
are small (see lower panel) the synchrotron-like radiation emitted
at the downstream edge of the cushion peaks at an energy
\eqb
h\nu_\gamma&\approx&0.7 mc^2
\left(r_{\rm c}/\alpha_{\rm f}\right)
\cos^3\delta_2
\label{gammaestimate}
\eqe
where $\alpha_{\rm f}$ is the fine-structure constant.

The lower panel of Fig.~\ref{reflect} plots the 
the QED parameter $\eta$, the ratio of the electric field
seen in the electron or positron rest frame to the critical 
field $E_{\rm c}=m^2c^3/e\hbar$ at the edge of the pair front.
As $\eta$ approaches unity, quantum effects such as electron
recoil on emitting a photon 
and pair production begin to become important.
Three curves are shown. 
At low intensities, where the number of pairs that
can be contained in a stationary cushion is relatively large, 
the solid red line (depicting $\eta$
in the presence of a maximal pair cushion) 
lies well below the 
dashed red line (depicting $\eta$ in 
the absence of a pair cushion.) At intensities
close to $10^{24}\wsqcm$, the reduction is less marked, being roughly 
$15\%$. Both of these curves lie below that predicted for counter-propagating
vacuum waves of intensity $I^+$
\cite{bellkirk08}, shown as a blue dashed line. This is because of 
the recoil of the target, 
which effectively reduces the pressure and frequency of the 
incident laser pulse \cite{ridgersetal12}. 
Assuming $a_0\gg1$, we find
\eqb
\eta_{\rm pairs}&=&\left(\hbar\omega/mc^2\right)a_0^2\cos^2\delta_2
\label{etapairs}
\eqe
At low intensity, $\eta_{\rm pairs}\propto a_0^2\propto \sqrt{I^+}$, but 
the factor $\cos^2\delta$, arising from the 
phase shift between the force on the charged fluids and the 
electric field which is brought about by radiation reaction, causes 
$\eta$ to rise less rapidly with laser intensity when 
$I^+>3\times10^{23}\wsqcm$ \cite{bellkirk08}. 

\section{Discussion}

The solutions presented above fulfil the fully nonlinear coupled fluid
and Maxwell equations including radiation reaction. Several studies
have treated classical radiation reaction in the contexts of the
hole-boring scenario and of counter-propagating laser beams
\cite{bulanovetal11,lehmannspatschek12,schlegeltikhonchuk12}, but
these neglect the influence of the radiating particles on the laser
fields.  Classical radiation reaction terms have been incorporated in
particle-in-cell simulation codes, which, in principle, treat the
fields self-consistently \cite{naumovaetal09,capdessusetal12,ceruttietal13}.
However, the solutions we find are analytical, in the sense that they
can be reduced to the 
quadratures (\ref{deltaprimeeq2}) and (\ref{phiintegral}).  This makes them 
a useful and flexible tool for the interpretation of 
both simulations and experiments. 

The unique aspect of these solutions is the inclusion of classical
radiation reaction. This force, acting on a single
charged (relativistic) particle moving in the coherent laser field, 
is approximately
anti-parallel to the particle speed in the lab.\ frame. The energy 
dissipated is
carried off primarily as short-wavelength photons. 
We assume the fluids in our treatment
consist of electrons and positrons that radiate independently of each
other, which is reasonable provided the wavelength of the radiated
photons in the particle rest-frame is small compared to the
inter-particle spacing in that frame. For photons of frequency
$a_0^3\omega_{\rm lab}$ close to the peak of the synchrotron-like
spectrum and for a pair plasma at the critical density, this
ratio can be estimated as
\eqb
\varepsilon&=&\frac{c/\left(a_0^2\omega_{\rm lab}\right)}{ \left(4\pi
    e^2/m\omega_{\rm lab}^2\right)^{1/3}}
\nonumber\\
&=&\left(a_0^3 r_{\rm c}\right)^{-1/3} 
\nonumber\\
&=&
6\times10^{-4}D^{-1/3}\left(1+R'\right)^{-1}
\lambda_{\mu{\rm m}}^{-5/3}
\left(I^+/10^{24}\wsqcm\right)^{-1}
\eqe 
so that our approach is justified for optical lasers with
$I^+>10^{21}\wsqcm$.

In this regime, classical radiation reaction introduces an effective
friction term into the equations of motion for the fluids.  For the
one-dimensional problem considered here, this force lies in the
$y$--$z$ plane and has no component in the direction of laser
propagation. As a result, the pair cushion remains in place provided
the pressure in the electromagnetic fields of the laser,
$\left(E^2+B^2\right)/8\pi$, remains constant. On the other hand,
energy conservation requires that the Poynting flux in the cushion
decrease monotonically towards the target. This is achieved by
allowing $\left|B\right|$ to decrease, reaching zero at the front edge
of the cushion, which, therefore, matches to a node of the magnetic
field in the standing wave that separates it from the target.

As noted in section~\ref{twofluidmodel}, the effective friction force
does not depend on which of the Lorentz-Abraham-Dirac or the
Landau-Lifshitz formulations of the radiation reaction term is
used. The underlying reason is that we have treated only stationary
solutions of the equations. An investigation of the stability
properties of the solutions, on the other hand, can be expected to
reveal a difference. 

However, our results should also be modified by quantum effects. When
$\eta_{\rm pairs}\sim1$, the characteristic energy of the radiated
photons is comparable to that of the radiating particle. In this case
losses become a discrete process, and cannot be represented by a
\lq\lq smooth\rq\rq\ friction term. As is well-known in the 
analogous case
of betatron oscillations,
this is likely to have a strong
influence on the stability properties of the solutions~\cite{ternov95}. 
But, even
before the discrete nature becomes pronounced, quantum effects
significantly reduce the time-averaged energy loss rate.  For example,
when $\eta\approx 1/10$, an emitted photon takes off only 10\% of the
particle energy, but the time-averaged energy loss rate is reduced by
a factor of 1/3. This reduction can be accounted for in an approximate
manner by modifying the radiation reaction term \cite{kirketal09}. 
Using this approach, it would be possible to improve the 
rough estimate (\ref{gammaestimate}) of the energy spectrum of
the emitted gamma-rays, although this would necessitate a
more elaborate numerical solution.

In common with all analytical solutions, those we present above 
suffer from several limitations. 
For low laser intensities, the linear extent of 
the maximal cushion becomes large 
(see Fig.~\ref{reflect}), so that the stationary solution 
can be realized only for a rather long
incident pulse. Also,
it is not clear that a significant number
of pairs will be available to form a cushion
in a realistic experimental configuration, particularly at low 
laser intensity.  
At high intensity, on the other hand, the
underlying hole-boring scenario itself is in doubt.
Piston oscillations can be become pronounced, and hot electrons may leak 
from the target into the vacuum gap 
\cite{robinsonetal09,schlegeletal09,naumovaetal09}.
Eventually, even solid targets 
become relativistically under-dense and are unable to reflect the
incident pulse. In Fig.~\ref{reflect} we 
have implicitly adopted the conventionally estimated 
threshold for this effect, although possible departures from it
have been discussed in the 
literature \cite{cattanietal00,wengetal12,siminosetal12}. 

\section{Conclusions}
We present 
solutions to the coupled set of Maxwell's equations
and those for two cold, charged, relativistic fluids
(pair plasma), including classical 
radiation reaction. The pair plasma reaches critical density and
is bounded by regions containing
vacuum fields: an incident laser and its reflected beam on one side,
and a standing vacuum wave separating the pair plasma or 
\lq\lq cushion\rq\rq\ from an over-dense target on the other. 
The solutions have two main properties, both of 
which are shown in 
Fig.~\ref{reflect}: First, the pair cushion forms an
efficient device for converting the energy flux in the laser into high
energy photons, as is evident from the substantial reduction in the
reflectivity.  Second, the laser
intensity at which quantum effects become important is increased
somewhat, as can be seen from the difference between 
$\eta_{\rm pairs}$ and $\eta_{\rm vac}$. 

However, it is not possible, using our calculations, to make
a reliable estimate of the threshold for the onset of a nonlinear pair
cascade, since important effects such as \lq\lq straggling\rq\rq\
\cite{duclousetal11} are not considered.

\ack 
ARB and CPR thank the UK Engineering and Physical Sciences Research 
Council for support under grant EP/G055165/1.

\section*{References}

\begin{thebibliography}{10}

\bibitem{nerushetal11}
E.~N. {Nerush}, I.~Y. {Kostyukov}, A.~M. {Fedotov}, N.~B. {Narozhny}, N.~V.
  {Elkina}, and H.~{Ruhl}.
\newblock {Laser Field Absorption in Self-Generated Electron-Positron Pair
  Plasma}.
\newblock {\em Physical Review Letters}, 106(3):035001, January 2011.

\bibitem{ridgersetal12}
C.~P. {Ridgers}, C.~S. {Brady}, R.~{Duclous}, J.~G. {Kirk}, K.~{Bennett}, T.~D.
  {Arber}, A.~P.~L. {Robinson}, and A.~R. {Bell}.
\newblock {Dense Electron-Positron Plasmas and Ultraintense {$\gamma$} rays
  from Laser-Irradiated Solids}.
\newblock {\em Physical Review Letters}, 108(16):165006, April 2012.

\bibitem{bradyetal12}
C.~S. {Brady}, C.~P. {Ridgers}, T.~D. {Arber}, A.~R. {Bell}, and J.~G. {Kirk}.
\newblock {Laser Absorption in Relativistically Underdense Plasmas by
  Synchrotron Radiation}.
\newblock {\em Physical Review Letters}, 109(24):245006, December 2012.

\bibitem{bellkirk08}
A.~R. {Bell} and J.~G. {Kirk}.
\newblock {Possibility of Prolific Pair Production with High-Power Lasers}.
\newblock {\em Physical Review Letters}, 101(20):200403--+, November 2008.

\bibitem{fedotovetal10}
A.~M. {Fedotov}, N.~B. {Narozhny}, G.~{Mourou}, and G.~{Korn}.
\newblock {Limitations on the Attainable Intensity of High Power Lasers}.
\newblock {\em Physical Review Letters}, 105(8):080402, August 2010.

\bibitem{robinsonetal09}
A.~P.~L. {Robinson}, P.~{Gibbon}, M.~{Zepf}, S.~{Kar}, R.~G. {Evans}, and
  C.~{Bellei}.
\newblock {Relativistically correct hole-boring and ion acceleration by
  circularly polarized laser pulses}.
\newblock {\em Plasma Physics and Controlled Fusion}, 51(2):024004, February
  2009.

\bibitem{schlegeletal09}
T.~{Schlegel}, N.~{Naumova}, V.~T. {Tikhonchuk}, C.~{Labaune}, I.~V. {Sokolov},
  and G.~{Mourou}.
\newblock {Relativistic laser piston model: Ponderomotive ion acceleration in
  dense plasmas using ultraintense laser pulses}.
\newblock {\em Physics of Plasmas}, 16(8):083103, August 2009.

\bibitem{naumovaetal09}
N.~{Naumova}, T.~{Schlegel}, V.~T. {Tikhonchuk}, C.~{Labaune}, I.~V. {Sokolov},
  and G.~{Mourou}.
\newblock {Hole Boring in a DT Pellet and Fast-Ion Ignition with Ultraintense
  Laser Pulses}.
\newblock {\em Physical Review Letters}, 102(2):025002, January 2009.

\bibitem{dipiazzaetal12}
A.~{Di Piazza}, C.~{M{\"u}ller}, K.~Z. {Hatsagortsyan}, and C.~H. {Keitel}.
\newblock {Extremely high-intensity laser interactions with fundamental quantum
  systems}.
\newblock {\em Reviews of Modern Physics}, 84:1177--1228, July 2012.

\bibitem{marburgertooper75}
J.~H. {Marburger} and R.~F. {Tooper}.
\newblock {Nonlinear optical standing waves in overdense plasmas}.
\newblock {\em Physical Review Letters}, 35:1001--1004, October 1975.

\bibitem{bulanovetal11}
S.~V. {Bulanov}, T.~Z. {Esirkepov}, M.~{Kando}, J.~K. {Koga}, and S.~S.
  {Bulanov}.
\newblock {Lorentz-Abraham-Dirac versus Landau-Lifshitz radiation friction
  force in the ultrarelativistic electron interaction with electromagnetic wave
  (exact solutions)}.
\newblock {\em \pre}, 84(5):056605, November 2011.

\bibitem{lehmannspatschek12}
G.~{Lehmann} and K.~H. {Spatschek}.
\newblock {Phase-space contraction and attractors for ultrarelativistic
  electrons}.
\newblock {\em \pre}, 85(5):056412, May 2012.

\bibitem{schlegeltikhonchuk12}
T.~{Schlegel} and V.~T. {Tikhonchuk}.
\newblock {Classical radiation effects on relativistic electrons in
  ultraintense laser fields with circular polarization}.
\newblock {\em New Journal of Physics}, 14(7):073034, July 2012.

\bibitem{capdessusetal12}
R.~{Capdessus}, E.~{d'Humi{\`e}res}, and V.~T. {Tikhonchuk}.
\newblock {Modeling of radiation losses in ultrahigh power laser-matter
  interaction}.
\newblock {\em \pre}, 86(3):036401, September 2012.

\bibitem{ceruttietal13}
B.~{Cerutti}, G.~R. {Werner}, D.~A. {Uzdensky}, and M.~C. {Begelman}.
\newblock {Simulations of Particle Acceleration beyond the Classical
  Synchrotron Burnoff Limit in Magnetic Reconnection: An Explanation of the
  Crab Flares}.
\newblock {\em \apj}, 770:147, June 2013.

\bibitem{ternov95}
I.~M. {Ternov}.
\newblock {REVIEWS OF TOPICAL PROBLEMS: Synchrotron radiation}.
\newblock {\em Physics Uspekhi}, 38:409--434, April 1995.

\bibitem{kirketal09}
J.~G. {Kirk}, A.~R. {Bell}, and I.~{Arka}.
\newblock {Pair production in counter-propagating laser beams}.
\newblock {\em Plasma Physics and Controlled Fusion}, 51(8):085008, August
  2009.

\bibitem{cattanietal00}
F.~{Cattani}, A.~{Kim}, D.~{Anderson}, and M.~{Lisak}.
\newblock {Threshold of induced transparency in the relativistic interaction of
  an electromagnetic wave with overdense plasmas}.
\newblock {\em \pre}, 62:1234--1237, July 2000.

\bibitem{wengetal12}
S.~M. {Weng}, M.~{Murakami}, P.~{Mulser}, and Z.~M. {Sheng}.
\newblock {Ultra-intense laser pulse propagation in plasmas: from classic
  hole-boring to incomplete hole-boring with relativistic transparency}.
\newblock {\em New Journal of Physics}, 14(6):063026, June 2012.

\bibitem{siminosetal12}
E.~{Siminos}, M.~{Grech}, S.~{Skupin}, T.~{Schlegel}, and V.~T. {Tikhonchuk}.
\newblock {Effect of electron heating on self-induced transparency in
  relativistic-intensity laser-plasma interactions}.
\newblock {\em \pre}, 86(5):056404, November 2012.

\bibitem{duclousetal11}
R.~{Duclous}, J.~G. {Kirk}, and A.~R. {Bell}.
\newblock {Monte Carlo calculations of pair production in high-intensity
  laser-plasma interactions}.
\newblock {\em Plasma Physics and Controlled Fusion}, 53(1):015009, January
  2011.

\end{thebibliography}

\end{document}